\documentclass{jetplA}
\twocolumn

\usepackage{hyperref}    

\lat

\title{On Phase Transitions near Black Holes}

\rtitle{On phase transitions near black holes}

\sodtitle{On Phase Transitions near Black Holes}

\author{\bf A.\,A. Grib$^{a,}$\/\thanks[1]{e-mail: andrei\_grib@mail.ru},
Yu.\,V. Pavlov$^{b,c,}$\/\thanks[2]{e-mail: yuri.pavlov@mail.ru}}

\rauthor{A.\,A. Grib, Yu.\,V. Pavlov}

\sodauthor{Grib, Pavlov}

\address{$^a$\,Hertzen University, St. Petersburg, 191186 Russia\\~\\
$^b$\,Institute of Problems of Mechanical Engineering, Russian Academy of
Sciences, St. Petersburg, 199178 Russia\\~\\
$^c$\,Lobachevsky Institute of Mathematics and Mechanics, Kazan Federal
University, Kazan, 420008 Russia}

\dates{September 9, 2022}{*}

\abstract{It has been shown that temperatures near the horizon of rotating
black holes can be about the phase transition temperature in the Standard Model
with the Higgs boson.
    The distance from the horizon and gravitational and electromagnetic
radiation emitted in collisions between particles have been numerically
estimated.}


\begin{document}

\maketitle

\section{Introduction}
\label{secIn}

    A quark-gluon plasma with a temperature of $4 \times 10^{12}$\,K
     was obtained in collisions of gold ions in the relativistic heavy ion
collider at the Brookhaven National Laboratory (United States) in 2010.
    The production of the quark–gluon plasma with a temperature of
$4 \times 10^{12}$\,K in collisions of lead ions at energies about several
TeV per colliding nucleon at the Large Hadron Collider (LHC) was reported
in 2012.
    However, the study of collisions of particles near rotating black
holes~\cite{BSW}--\cite{Zaslavskii21} shows that there is a natural
supercollider where collision energies are much higher than those reached
at the LHC.
    Consequently, the quark–gluon plasma produced in such collisions should
have very high temperatures.
    It is of interest whether these temperatures can be as high as the phase
transition temperature in the quark–gluon plasma, in particular, in the
theory of the electroweak interaction.
    At this temperature, vacuum is changed; as a result, the vacuum average
of the Higgs field and thereby the quark masses vanish and the properties of
the quark–gluon plasma change significantly.
    The aim of this work is to reveal the conditions under which this
phenomenon occurs, including the distance from the horizon of events of
a black hole.
    The phase transition in electroweak interactions in cosmology was
previously discussed in~\cite{KirzhnitsLinde74,KirzhnitsLinde76,Weinberg74}
and other works.

\section{Phase transitions in the early Universe}
\label{secl}

    In the extrapolation of the standard cosmological model to times close
to the Big Bang time, very high temperatures can theoretically be reached at
which the following transitions between states of space matter can occur,
usually called cosmological phase transitions.

    (i) Transition between the quark.gluon plasma and hadrons at energies
$E \sim 200$ MeV.
    The corresponding temperature $T=E/k_B \approx 10^{12}$\,K,
where $k_B= 1.380649\cdot 10^{-23}$\,J/K is the Boltzmann constant,
can exist in the expanding Universe at a time of about
$10^{-6}$\,s after the Big Bang.

    (ii) The electroweak phase transition at energies
$E_W \approx 100$\,GeV.
    The corresponding temperature $T_W \approx 10^{15}$\,K can exist at a time
of about $10^{-12}$\,s after the Big Bang.

    (iii) The grand unification phase transition at energies
$E_{\rm GUT} \approx10^{16}$\,GeV.
    The corresponding temperature
$T_{\rm GUT}=E_{\rm GUT}/k_B \approx 10^{29}$\,K can hardly be reached in the
early Universe~\cite{GorbunovRubakov} according to models with an inflation
stage in which the heating temperature is much lower than~$T_{\rm GUT}$.
    According to models with an radiation dominant stage, the
temperature~$T_{\rm GUT}$ can be reached in the early Universe at times
about $10^{-38}$\,s.

    The possible phase transitions in the early Universe lead to the known
problem of the cosmological constant~\cite{GorbunovRubakov}.
    Indeed, the energy density of vacuum corresponding to the mentioned phase
transitions is many orders of magnitude higher than the value corresponding
to the observed cosmological constant.
    To explain this, the fine tuning of the bare cosmological constant
is necessary.

\section{Hawking temperature near the horizon of a nonrotating black hole}
\label{sec2p}

    High temperatures and regions of possible phase transitions can occur
near the horizon of black holes owing to the thermal emission of black holes
discovered by Hawking~\cite{Hawking74,Hawking75}.

    The metric of a nonrotating black hole can be represented in the form
    \begin{equation}
d s^2 = \left( 1 - \frac{r_g}{r} \right) c^2 d t^2 -
\frac{d r^2}{ 1 - \frac{r_g}{r} } -r^2 d \Omega^2 .
\label{m1}
\end{equation}
    Here $r_g = 2 GM /c^2$, where $G$ is the gravitational constant and
$M$ is the mass of the black hole; $c$ is the speed of light in vacuum; and
$d \Omega^2 = d \theta^2 + \sin^2 \theta \, d \varphi^2 $.
    The Hawking temperature at infinite distance from the Schwarzschild black
hole is~\cite{GMM}
    \begin{equation}
T_H = \frac{\hbar c^3}{8 \pi k_B G M} \approx 6.169\cdot 10^{-8} K \cdot
\frac{M_\odot}{M},
\label{m2}
\end{equation}
    where $ \hbar = 1.05457 \cdot 10^{-34}$\,J$\cdot$s is the reduced
Planck constant and $M_\odot$ is the mass of the Sun.

    As known~\cite{Tolman}
    \begin{equation}
T \sqrt{g_{00} } = {\rm const}.
\label{m3}
\end{equation}
    in the gravitational field in thermodynamic equilibrium.
    Consequently, the Hawking temperature of the nonrotating black hole
at points with the radial coordinate~$r$ is given by the expression
    \begin{eqnarray}
T(r) &=&  T_H / \sqrt{1 - \frac{r_g}{r} }= \frac{\hbar c}{4 \pi k_B r_g}
\sqrt{\frac{r}{\Delta r}}
    \nonumber \\
&\approx& 6.169\cdot 10^{-8} K \cdot
\frac{M_\odot}{M} \sqrt{\frac{r}{\Delta r}},
\label{m4}
\end{eqnarray}
    where $\Delta r = r - r_g$.
    For radial coordinates $r\approx r_g$ at which the temperature of the
Hawking radiation is $T$,
    \begin{equation}
\Delta r \approx \frac{\hbar^2 c^4}{32 \pi^2 k_B^2 G M T^2} \approx
\frac{1.12 \cdot 10^{-11}}{T^2} \frac{M_\odot}{M}\ ({\text m}).
\label{m5}
\end{equation}
    Here, the temperature is given in kelvins and the distance is in meters.
    The substitution of the quark-gluon phase transition temperature and
$M = M_\odot$ into Eq.~(\ref{m5}) gives $\Delta r \approx 10^{-35}$\,m,
    which is about the Planck length
$l_{Pl} = \sqrt{\hbar G/c^3} \approx 1.616 \cdot 10^{-35}$\,m.
    This distance for black holes with masses larger than the mass of the
Sun is even smaller.

    It is noteworthy that the radial coordinate $r$ is not identical to
the physical distance, which can be defined only locally in the curved
spacetime~\cite{LL_II}.
    To illustrate the meaning of the radial coordinate, we present the radial
equation of timelike geodesics in the Schwarzschild field~(\ref{m1})
(see~\cite{Chandrasekhar}, Section 19]):
    \begin{equation}
\left( \frac{d r}{c\, d \tau}\right)^2 = \varepsilon^2 - \left( 1 - \frac{r_g}{r}
\right) \left( 1 + \frac{L^2}{(mcr)^2} \right),
\label{GeSch}
\end{equation}
    where $ \tau$, $\varepsilon = E/ (m c^2) $, and $ L $
are the proper time, specific energy, and angular momentum of the moving
particle with the mass~$m$.
    According to Eq.~(\ref{GeSch}),
    \begin{equation}
d \tau = \frac{ d r }{ c \sqrt{ \varepsilon^2 - \left( 1 - \frac{r_g}{r} \right)
\left( 1 + \frac{L^2}{(mcr)^2} \right)} } .
\label{GeSch2}
\end{equation}
    According to Eq.~(\ref{GeSch2}) at $r \to r_g$, the residence time of
the particle with a fixed angular momentum in the region~$d r$ is
$ \approx dr/(\varepsilon c)$.

    The total energy of the particle incident from a region far from the event
horizon cannot be much lower than $mc^2$, and, therefore, the residence time
in the indicated region does not exceed $dr/c$ in order of magnitude.
    Thus, the above estimates of $ \Delta r$ for the quark-gluon phase
transition show that matter outside the event horizon can be in such state for
a time about the Planck time, which is physically unobservable.
    Consequently, phase transitions, even quark-gluon ones, near the horizon
of black holes at the Hawking temperature cannot be observed.

\section{Temperature reached in collisions near the horizon of
an extreme rotating black hole}
\label{secTBH}

    The Kerr metric of a rotating black hole~\cite{Kerr63} in the
Boyer–Lindquist coordinates~\cite{BoyerLindquist67} has the form
    \begin{equation}
d s^2 = \frac{\rho^2 \Delta}{\Sigma^2}\,c^2 d t^2 -
\frac{\sin^2\! \theta}{\rho^2} \Sigma^2 \, ( d \varphi - \omega d t)^2
\label{Kerr}
- \frac{ \rho^2}{\Delta}\, d r^2 - \rho^2 d \theta^2 ,
\end{equation}
    where
    \begin{equation} \label{Delta}
\rho^2 = r^2 + \frac{a^2}{c^2} \cos^2 \! \theta, \ \ \ \ \
\Delta = r^2 - \frac{2 G M r}{c^2} + \frac{a^2}{c^2},
\end{equation}
    \begin{equation} \label{Sigma}
\Sigma^2 = \left( r^2 + \frac{a^2}{c^2} \right)^2 -
\frac{a^2}{c^2} \sin^2\! \theta\, \Delta , \ \ \ \
\omega = \frac{2 G M r a}{\Sigma^2 c^2} ,
\end{equation}
    $M$ and $ aM $ are the mass and angular momentum of the black hole,
respectively.
    We accept that  ${0 \le a \le G M/c }$.
    The event horizon of the Kerr black hole has the radial coordinate
    \begin{equation}
r = r_H \equiv \frac{G}{c^2} \left( M + \sqrt{M^2 - \left( \frac{a c}{G}
\right)^2} \right) .
\label{Hor}
\end{equation}

    According to~\cite{BSW} the squared energy of collision of two particles
with the mass~$m$ with the angular momenta $L_1$ and $L_2$ in
the center-of-mass system, which are nonrelativistic at infinity and are
freely incident on a black hole with the angular momentum~$aM$, is given
by the expression
    \begin{eqnarray}
\frac{E_{\rm c.m}^2}{m^2 c^4} = \frac{2}{x(x^2-2 x+A^2)} \Biggl[
2 A^2 (1+ x) \ \ \ \ \ \nonumber \\   \label{BSW}
-\, 2 A (l_1+l_2) - l_1 l_2 (x-2) + 2(x-1) x^2 \ \ \ \ \\
-\, \sqrt{2(A\!-\!l_2)^2 \!- l_2^2 x + 2 x^2 } \sqrt{2(A \! -\! l_1)^2
\!- l_1^2 x + 2 x^2 }
\Biggr],  \!\!\!\! \nonumber
\end{eqnarray}
    where $x= r c^2/ G M$, $l= L c/ G m M$, and $A= a c/ G M$.
    For the extreme rotating black hole, $A=1$, and the event horizon
corresponds to $x=x_H=1$.
    For an incident particle to reach the event horizon, its angular
momentum should not be high in absolute value.
    In particular, to reach the event horizon of the black hole, the particle,
which is nonrelativistic at infinity ($E/ m c^2 = 1$) and is incident in
the equatorial plane, should have the angular momentum in the range
    \begin{eqnarray}
-2 [1+\sqrt{1+A} ] \le l \le 2 [1+\sqrt{1-A} ].
\label{Llim}
\end{eqnarray}
    The energy of collision of two particles with the angular momenta
$l_1=4$ and $l_2=-4$ near the event horizon of the nonrotating ($A=0$)
black hole is $2 \sqrt{5} m c^2$.
    The maximum achievable energy of collision in the center-of-mass
system increases with the velocity of rotation of
the black hole~\cite{GribPavlov2011}.

    The particle with the maximum possible angular momentum $l_2=2$
(critical particle) freely incident on the extreme rotating black hole will be
twisted on the event horizon for an infinitely long proper time.
    The energy of collision of this critical particle with another particle
with the angular momentum $l_1$ in the interval of ($-2(1+\sqrt{2}) , \, 2$)
near the event horizon can be unlimitedly high
(Banados–Silk–West resonance~\cite{BSW}):
    \begin{equation}
\frac{E_{\rm c.m}^2}{m^2 c^4} = \frac{4 }{x(x\!-\!1)} \biggl(
1 -\! l_1 + x^2 \!-\! \sqrt{(1 \!-\! l_1)^2 \!- l_1^2 x/2 +\! x^2 } \biggr).
\label{BSW1}
\end{equation}
    The substitution of this energy into
the formula $T=(E_{\rm c.m}-2 mc^2)/k_B$ for the temperature yields
    \begin{eqnarray}
T = \frac{2 m c^2}{k_B} \Biggl[ \sqrt{ \frac{ 1 -\! l_1 + x^2 \!-\!
\sqrt{(1 \!-\! l_1)^2 \!- l_1^2 x/2 +\! x^2 }} {x(x- 1)}} \nonumber \\
-\,1 \Biggr] \approx 1.083 \frac{m c^2}{k_B} \sqrt{ \frac{2-l_1}{ x-1}},
\ \ \ x \to 1 . \hspace{4mm}
\label{BSW4}
\end{eqnarray}
     This temperature is reached at the distance
    \begin{equation}
r - r_H \approx 1.17 \cdot (2-l_1) r_H \left( \frac{m c^2}{k_B T} \right)^2,
\ \ \ r \to r_H .
\label{BSW5}
\end{equation}
     For the mass $m$ about the proton mass and $l_1=0$, the electroweak
temperature near the extreme rotating black hole is reached at the distance
    \begin{equation}
r - r_H \approx 2 \cdot 10^{-4} r_H .
\label{BSW6}
\end{equation}
    This distance for black holes with the mass of the Sun is tens of
centimeters and, in contrast to the Hawking temperature, is acceptable for
the occurrence of a phase transition.

    To estimate the corresponding temperature far from the black hole,
we use the equation for the time component of the timelike
geodesic~\cite{Chandrasekhar}
    \begin{equation}
\rho^2 \frac{d t}{d \tau} = \frac{1}{\Delta } \left( \Sigma^2 \varepsilon
- 2 \left( \frac{GM}{c^2} \right)^2 r l \frac{a}{c} \right).
\label{geod0}
\end{equation}
    For the particle with $l< 2 \varepsilon r_H c/a$
(noncritical particle) freely incident near the horizon of the rotating black
hole, the dilation of the proper time $\tau$ compared to the time at
infinity $t$ is
    \begin{equation}
\frac{d t}{d \tau} \sim \frac{2 \varepsilon (2 \varepsilon - l)}{(x-1)^2
(1+ \cos^2 \theta) }, \ \ x \to 1 .
\label{geod1}
\end{equation}
    The dilation of the proper time for the critical particle
with $l= l_H= 2 \varepsilon r_H c/a$ is
    \begin{equation}
\frac{d t}{d \tau} \sim \frac{4 \varepsilon}{(x-1)
(1+ \cos^2 \theta) }, \ \ x \to 1 .
\label{geod2}
\end{equation}

    Constraints on the possible specific energy $\varepsilon $ and
the projection of the angular momentum~$l$ on the axis of rotation of the
black hole at a given radial coordinate~$r$ can be obtained from equations
for the radial and angular components of geodesics.
    We consider only the motion in the equatorial plane with $\theta = \pi/2$.
    In this case, the equation for the radial component of
a geodesic has the form~\cite{Chandrasekhar}
    \begin{equation}
\frac{\rho^2 }{c} \frac{d r}{d \tau} = \pm \sqrt{R},
\label{geodre}
\end{equation}
    \begin{equation}
R = \Sigma^2 \varepsilon^2 \!- l r \left( \frac{GM}{c^2} \right)^{\!2} \!\!
\left( 4 \varepsilon \frac{a}{c} + l \left( r - 2 \frac{GM}{c^2} \right)
\right) \! -\! r^2 \Delta .
\label{geodR}
\end{equation}
    For the critical particle, the condition $R\ge 0$ gives the constraint
$\varepsilon \ge \sqrt{x/(x+2)}$; hence, the time dilation is
    \begin{equation}
\frac{d t}{d \tau} \gtrsim \frac{4}{\sqrt{3} (x-1)}, \ \ x\to 1 .
\label{geodZVK}
\end{equation}
    The specific energy of noncritical particles incident on the extreme
black hole can be low.
    In particular, the condition $R\ge 0$ in the case  $l= 0$ yields
$\varepsilon \ge \varepsilon_{\rm min}$, where
    \begin{equation}
\varepsilon_{\rm min} = \frac{(x-1) \sqrt{x} }{ \sqrt{ x^3 + x +2 } } \sim
\frac{x-1}{2}, \ \ x \to 1.
\label{geod3}
\end{equation}
    In the limiting case $\varepsilon_{\rm min}$, according to Eq.~(\ref{geod0}),
the time dilation is
    \begin{equation}
\frac{d t}{d \tau} = \frac{\sqrt{x^2 + 1 + 2/x}}{x-1}
\sim \frac{2}{x-1}, \ \ x \to 1 .
\label{geod4}
\end{equation}

    The collision of particles near the event horizon of the black hole with
the production of the quark-gluon plasma can be accompanied by the production
of particles with various specific energies and momenta and the subsequent
emission of photons at their collisions.
    According to Eqs.~(\ref{geod1}), (\ref{geodZVK}), and (\ref{geod4}),
the energy of a collision occurring at the point approaching the event horizon
of the black hole $x\to 1$, as well as the local temperature of produced
particles, can unlimitedly increase, but the temperature observed far from
the black hole tends to zero owing to the redshift.

    Combining Eqs.~(\ref{BSW4}) and (\ref{geod4}), we obtain the following
approximate expression for the temperature observed far from the black hole as
a function of the radial coordinate~$x$ of the collision point:
    \begin{eqnarray}
T(x) \approx \frac{m c^2}{k_B} \cdot \frac{2 (x-1)}{\sqrt{x^2 +x +2/x}}
\hspace{10mm} \nonumber \\
\times \Biggl[ \sqrt{ \frac{ 1 -\! l_1 + x^2 \!-\!
\sqrt{(1 \!-\! l_1)^2 \!- l_1^2 x/2 +\! x^2 }} {x(x- 1)}} - 1 \Biggr].
\label{Tx}
\end{eqnarray}
    The maximum $T(x)$ value at $l_1=-4$ is $ \approx 0.534 m c^2 / k_B$ and
is reached at $x \approx 1.744 $.

    The corresponding expression in the case of the collision of particles
with $l_1=-4$, $l_2 = 4$, and $E_{1,2} = mc^2$ incident on the Schwarzschild
black hole has the form
    \begin{eqnarray}
T(x) \approx \frac{m c^2}{k_B} \cdot \frac{32}{x \left(x + \sqrt{x^2 +16}
\right)} \sqrt \frac{x-2}{x} .
\label{TxSH}
\end{eqnarray}
    In this case, the maximum $T(x)$ value is $ \approx 0.803 m c^2 / k_B$
and is reached at $x \approx 2.645 $.

    Thus, the temperature observed far from high-energy collisions near
the event horizon of the black hole rotating at any velocity is in order of
magnitude no more than several tenths of $m c^2/ k_B $.
    The local temperature near the collision point near the extreme rotating
black hole can be unlimitedly high.
    Because of the conservation laws, the total energy emitted from the
black hole cannot be higher than the sum of the energies of the colliding
particles ($2 mc^2$ for the collision of particles with the same mass~$m$,
which are nonrelativistic at infinity, if the Penrose effect for rotating
black holes is disregarded).

    It is remarkable that the extreme rotating black hole is no longer
extreme after the incidence of any particle on it~\cite{GribPavlov2016}.
    Consequently, extreme black holes can hardly exist in nature.
    The extreme angular momentum of the black hole achievable upon the
accretion of matter on it is estimated in~\cite{Thorne74} as $A = 0.998$.
    At this angular momentum of the black hole, the maximum energy of
a collision in the center-of-mass system of freely incident particles
nonrelativistic at infinity is only $E_{\rm c.m.} \approx 19 mc^2 $.

\section{Multiple collisions near the horizon of nonextreme rotating black holes}
\label{secTnevbh}

    As shown in~\cite{GribPavlov2010,GribPavlov2011}, a superhigh center-of-mass
energy can be achieved in multiple collisions near nonextreme black holes.
    To reach the horizon, particles incident from infinity should have
an angular momentum low in absolute value.
    The angular momentum of one of the particles necessary for a high-energy
collision can be acquired either in multiple collisions or in the interaction
with the electromagnetic field of the accretion disk.
    The possible collision energy in the center-of-mass system of the particle
with the angular momentum~$l$ and the other particle with the same mass~$m$
can be estimated as~\cite{GribPavlov2010}
    \begin{equation}
E_{\rm c.m.}(r) \approx \frac{mc^2}{\sqrt{\delta}} \sqrt{
\frac{2(l_H-l)}{1 - \sqrt{1-A^2}}},
\label{Bbb1}
\end{equation}
    where $\delta$ is the parameter characterizing the region of radial
coordinates where the high-energy collision is possible:
    \begin{equation}
x \le x_H + \frac{\delta^2 (1-\sqrt{1-A^2})^2}{4 x_H \sqrt{1-A^2}}.
\label{Bbb2}
\end{equation}
    According to Eqs.~(\ref{Bbb1}) and (\ref{Bbb2}), the distance at which
the collision energy $E_{\rm c.m.}$ can be reached is given by the
expression
    \begin{equation}
x - x_H \approx \frac{(l_H-l)^2}{x_H \sqrt{1-A^2}}
\left( \frac{m c^2 }{E_{\rm c.m.}} \right)^4 .
\label{Bbb4}
\end{equation}
    In particular, for rotating black holes with $ A \approx 1 $ at $l=0$,
    \begin{equation}
r - r_H \approx \frac{2.8 \cdot r_H}{\sqrt{1-A}}
\left( \frac{m c^2 }{E_{\rm c.m.}} \right)^4 .
\label{Bbb6}
\end{equation}
    At the limiting value $A = 0.998$~\cite{Thorne74}, the electroweak
temperature in order of magnitude is possible at
$r - r_H \approx 6 \cdot 10^{-3} r_H $.
    This distance is about meters for black holes with the mass of the Sun
and can be thousands of kilometers for supermassive black holes.

    Thus, the quark–gluon and even electroweak phase transition temperatures
can be reached in multiple collisions of particles near rotating black holes.

\section{Collisions of macroscopic bodies}
\label{secMSSt}

    Collisions of macroscopic bodies near the event horizon of black holes are
possible only when such bodies reach the horizon region rather than being
destroyed by tidal forces of the gravitational field.
    Many processes of destruction of macroscopic objects (stars) by tidal
forces near supermassive black holes were observed in the SRG/eROSITA space
experiment~\cite{Sazonov21}.
    Tidal forces near the horizon decrease with an increase in the mass of
the black hole.
    We present below some estimates of the mass of black holes at which
macroscopic objects near the event horizon are not destroyed
(see also~\cite{FizikaKosmosa}, p.~772).
    We consider only the nonrotating black hole and radial tidal forces.
    According to the equation for the deviation of geodesics in the coordinates
associated with the center of mass of the incident body
(see~\cite{MTW}, Eq.~(32.24b)),
    \begin{equation}
\frac{D^2 \xi^r}{d \tau^2} = \frac{2GM }{r^3} \xi^r ,
\label{Bbb8}
\end{equation}
    where $\xi^r$ is the corresponding radial coordinate.
    For an estimate, we assume that a star or an incident planet is destroyed
if tidal forces acting on the points of the center of mass and the surface
exceed the force of attraction of the points of the surface to the center of
the incident body.
    The condition for the incidence of a uniform ball with the density~$\rho$
and radius~$R$ to the horizon has the form
    \begin{equation}
\frac{2GM}{r_g^3} R < \frac{G 4 \pi \rho R^3}{3 R^2} .
\label{Bbb9}
\end{equation}
    Therefore, the star or planet bound by the gravitational forces
is not destroyed upon incidence to the event horizon if the mass of
the black hole satisfies the inequality
    \begin{equation}
M > \frac{c^3}{4 G^{3/2}} \sqrt{\frac{3}{\pi \rho }}, \ \ \ \ \
\frac{M}{M_\odot} > 1.9 \cdot 10^8 \sqrt{\frac{\rho_W}{\rho}},
\label{Bbb10}
\end{equation}
    where $\rho_W = 10^3$\,kg/m$^3$.
    Thus, a collision of stars or planets with the relative velocity close to
the speed of light in vacuum is possible only near the event horizon
of supermassive black holes with the mass exceeding $10^8$ of the mass of
the Sun.
    Such black holes exist at the centers of many galaxies.
    The collision of compact objects with star masses near supermassive
black holes was considered in~\cite{HaradaKimura11b}.
    The authors of~\cite{Nayakshin12} showed that large clouds of comets,
asteroids, and stones should be formed around supermassive black holes.

    Let us estimate the mass of the black hole allowing incidence of
an undestroyed stone with a size of several centimeters/meters.
    The incident stone is destroyed if tidal forces exceed its tensile
strength for the radial direction or its compressive strength for the
polar and azimuthal directions.
    We consider only the case of breaking in the radial direction and take
the tensile strength of titanium alloy $\sigma = 10^9$\,Pa.
    The necessary condition for the body with the characteristic
size~$d$ and density~$\rho$ not to be destroyed has the form
    \begin{equation}
\frac{GM}{4r_g^3} \rho d^2 < \sigma
\label{Bbb11}
\end{equation}
    (see the derivation of Eq. (32.25a) in~\cite{MTW}).
    Substituting the expression~$r_g=2 GM/ c^2$ and the iron density
$\rho=7.87 \cdot 10^3$\,kg/m$^3$, we obtain
    \begin{equation}
M > \frac{c^3 d }{4 G} \sqrt{\frac{\rho}{2 \sigma }}, \ \ \ \ \
\frac{M}{M_\odot} > 10^2 d,
\label{Bbb12}
\end{equation}
    where $d$ is the distance in meters.
    Thus, collisions of $0.1$\,m stones with velocities close to the speed
of light in vacuum are possible near the horizon of black holes with stellar
masses.

    The collision energy per each particle in the body rather than the total
collision energy should be high to obtain high temperatures.
    The multiple collision mechanism is obviously inapplicable for macroscopic
bodies because any relativistic collision inevitably destroys bodies.
    Consequently, collisions with center-of-mass energies much higher than
$mc^2$ could be possible only near the horizon of extreme rotating black holes.

    The process of ultrarelativistic collisions of macroscopic bodies
expectedly has a complex structure.
    Such processes have not yet been observed in nature.
    Although the velocities of colliding neutron stars or merging black holes
that are detected by gravitational radiation bursts are close to the speed of
light in vacuum, the corresponding relativistic
factor $\gamma=1/\sqrt{1-v^2/c^2}$ only insignificantly exceeds unity.

    The kinetic energy of each particle in colliding ultrarelativistic
macroscopic bodies is much higher than the energy of electromagnetic bonds
between these particles in the bodies.
    Therefore, such bodies in the model of ultrarelativistic collisions
should be treated as colliding particle clouds.
    If the size $d$ of a cloud satisfies the inequality $\sigma n d \ll 1$,
where $\sigma$ is the effective scattering cross section and $n$ is the number
density of particles, only an insignificant number of particles in the cloud
are involved in the reaction and the colliding clouds pass through each other
without destruction.
    For an estimate, we take the area of the cross section of
a nucleus $\sigma = \pi r_n^2$ with the radius $ r_n \approx 10^{-15}$\,m
and the particle density of iron atoms $n=8.4 \cdot 10^{28}$\,m$^{-3}$.
    Then, colliding solid bodies with the characteristic
sizes $d \ge 30$ m will be destroyed.
    In this case, the collision of macroscopic bodies will lead to
ultrarelativistic pair collisions of nucleons constituting the bodies.

    Electromagnetic and gravitational radiation bursts should be expected
from ultrarelativistic collisions of macroscopic bodies.
    We assume that the collision of macroscopic bodies leads to pair collisions
of nucleons constituting these bodies and use estimates obtained
in~\cite{GribPavlov2019i}.
    Gravitational radiation from pair collisions of particles with the
mass~$m$ can be estimated by Eq. (10.4.23) from~\cite{Weinberg}.
    This radiation will be suppressed by a factor of $m^2/M_{Pl}^2$
(see Eq.~(11) in~\cite{GribPavlov2019i})),
where $M_{Pl}$ is the Planck mass.
    The energy of electromagnetic radiation in order of magnitude is
(see~\cite{GribPavlov2019i}, Eq.~(17))
    \begin{equation}
E_{\rm em} \approx \frac{e^2}{\hbar c} E_{\rm c.m.} ,
\label{Bbb13}
\end{equation}
    where $e$ is the elementary charge. The total intensity of electromagnetic
radiation from the collision of macroscopic bodies will obviously
be significant and can be observed even taking into account the redshift near
the horizon.
    In particular, the maximum center-of-mass energy of the collision
at points with a radial coordinate of $ r_H + 7 \cdot 10^5$\,km near
the horizon of the extreme rotating black hole with the mass $10^9 M_\odot$
can reach $100 mc^2$.
    In nucleon--nucleon collisions, this is the electroweak unification energy.
    The maximum total center-of-mass energy of the collision of two iron
asteroids with a diameter of 1 km with the corresponding $\gamma$-factor is
about $3 \cdot 10^{31}$\,J.
    The center-of-mass energy of electromagnetic radiation near the collision
point can be $2 \cdot 10^{29}$\,J at a power of about $10^{34}$\,W.
    Far from the black hole, an electromagnetic burst with an energy of
about $10^{25}$\,J at a power of about $10^{28}$\,W can be observed owing
to the redshift and time dilation in the gravitational field.

\section{Conclusions}
\label{secConcl}

    The Standard Model predicts the existence of the Higgs particle according
to the Higgs model based on the Goldstone model.
    As known~\cite{GribVac}, the Goldstone model is similar to the theory of
superfluidity and superconductivity and implies the existence of two vacua,
symmetric and asymmetric.
    The experimental discovery of the Higgs boson at the LHC allows the
possibility of a phase transition from one vacuum to the other at high
temperatures, as in the nonrelativistic quantum theory of many bodies, where
the ground state serves as vacuum.
    It has been shown in this work that such a phase transition in the
Standard Model is possible in multiple collisions near the horizon of
nonextremal rotating black holes (in their ergosphere).
    This phase transition is also possible in collisions of two macroscopic
bodies near extremal black holes in the presence of the Banados-Silk-West
resonance~\cite{BSW}.
    The emission of gravitational and electromagnetic waves accompanying the
phase transition has also been discussed. Gravitational radiation in these
collisions is insignificant, but electromagnetic radiation is fairly intense
and the corresponding burst can be observed on the Earth.

    This work was supported by the Russian Science Foundation
(grant no. 22-22-00112).



\begin{thebibliography}{99}

\bibitem{BSW}
M.\,Banados, J.\,Silk, and S.\,M.\,West,
{\em Kerr black holes as particle accelerators to arbitrarily high energy},
\href{https://doi.org/10.1103/PhysRevLett.103.111102}
{Phys. Rev. Lett. {\bf 103}, 111102 (2009)}.

\bibitem{GribPavlov2010}
A.\,A.\,Grib and Yu.\,V.\,Pavlov,
{\em On the collisions between particles in the vicinity of rotating black holes},
\href{https://doi.org/10.1134/S0021364010150014}
{JETP Lett. {\bf 92}, 125--129 (2010)}.

\bibitem{GribPavlov2011}
A.\,A.\,Grib and Yu.\,V.\, Pavlov,
{\em On particle collisions in the gravitational field of the Kerr black hole},
\href{https://doi.org/10.1016/j.astropartphys.2010.12.005}
{Astropart. Phys. {\bf 34}, 581--586 (2011)}.

\bibitem{Zaslavskii20}
O.\,B.\,Zaslavskii,
{\em Schwarzschild black hole as a particle accelerator},
\href{https://doi.org/10.1134/S0021364020050033}
{JETP Lett. {\bf 111}, 260--263 (2020)}; arXiv:1910.04068

\bibitem{Zaslavskii21}
O.\,B.\,Zaslavskii,
{\em Super-Penrose process for nonextremal black holes},
\href{https://doi.org/10.1134/S0021364021120043}
{JETP Lett. {\bf 113}, 757--762 (2021)}; arXiv:2103.02322

\bibitem{KirzhnitsLinde74}
D.\,A.\,Kirzhnits and A.\,D.\,Linde,
{\em Relativistic phase transition},
Sov. Phys. JETP. {\bf 40}, 628 (1975).

\bibitem{KirzhnitsLinde76}
D.\,A.\,Kirzhnits and A.\,D.\,Linde,
{\em Symmetry behavior in gauge theories},
\href{https://doi.org/10.1016/0003-4916(76)90279-7}
{Annals of Physics {\bf 101}, 195--238 (1976)}.

\bibitem{Weinberg74}
S.\,Weinberg,
{\em Gauge and global symmetries at high temperature},
\href{https://doi.org/10.1103/PhysRevD.9.3357}
{Phys. Rev.~D {\bf 9}, 3357--3378 (1974)}.

\bibitem{GorbunovRubakov}
D.\,S.\,Gorbunov and V.\,A.\,Rubakov,
{\em Introduction to the Theory of the Early Universe: The Hot Big Bang Theory}
(LENAND, Moscow, 2016; World Scientific, Singapore, 2011).

\bibitem{Hawking74}
S.\,W.\,Hawking,
{\em Black hole explosions?}
\href{https://doi.org/10.1038/248030a0}
{Nature {\bf 248}},
\href{https://doi.org/10.1038/248030a0}
{30--31 (1974)}.

\bibitem{Hawking75}
S.\,W.\,Hawking,
{\em Particle creation by black holes},
\href{https://doi.org/10.1007/BF02345020}
{Commun. Math. Phys. {\bf 43}, 199--220 (1975)}.

\bibitem{GMM}
A.\,A.\,Grib, S.\,G.\,Mamayev, and V.\,M.\,Mostepanenko,
{\em Vacuum Quantum Effects in Strong Fields}
(Energoatomizdat, Moscow, 1988; Friedmann Lab. Publ., St. Petersburg, 1994).

\bibitem{Tolman}
R.\,C.\,Tolman, {\em Relativity, Thermodinamics and Cosmology}
(Clarendon Press, Oxford, 1934).

\bibitem{LL_II}
L.\,D.\,Landau and E.\,M.\,Lifshitz, {\em Course of Theoretical Physics},
Vol.~2: {\em The Classical Theory of Fields}
(Pergamon, Oxford, 1983; Nauka, Moscow, 1988).

\bibitem{Chandrasekhar}
S.\,Chandrasekhar, {\em The Mathematical Theory of Black Holes}
(Oxford Univ. Press, Oxford, UK, 1983).

\bibitem{Kerr63}
R.\,P. Kerr,
{\em Gravitational field of a spinning mass as an example of algebraically
special metrics},
\href{http://dx.doi.org/10.1103/PhysRevLett.11.237}
{Phys. Rev.}
\href{http://dx.doi.org/10.1103/PhysRevLett.11.237}
{Lett. {\bf 11}, 237--238 (1963)}.

\bibitem{BoyerLindquist67}
R.\,H.\,Boyer and R.\,W.\,Lindquist,
{\em Maximal analytic extension of the Kerr metric},
\href{http://dx.doi.org/10.1063/1.1705193}
{J. Math. Phys. {\bf 8}},
\href{http://dx.doi.org/10.1063/1.1705193}
{265--281 (1967)}.

\bibitem{GribPavlov2016}
A.\,A.\,Grib and Yu.\,V.\,Pavlov,
{\em Black holes and particles with zero or negative energy},
\href{https://doi.org/10.1134/S0040577917020088}
{Theor. Math. Phys.}
\href{https://doi.org/10.1134/S0040577917020088}
{{\bf 190}, 268--278 (2017)}.

\bibitem{Thorne74}
K.\,S.\,Thorne,
{\em Disk-accretion onto a black hole. II. Evolution of the hole},
\href{https://doi.org/10.1086/152991}
{{Astrophys. J.} {\bf 191}, 507--519}
\href{https://doi.org/10.1086/152991}
{(1974)}.

\bibitem{Sazonov21}
S.\,Sazonov, M.\,Gilfanov, P.\,Medvedev, et al.,
{\em First tidal disruption events discovered by SRG/eROSITA: X-ray/optical
properties and X-ray luminosity function at $z < 0.6$},
\href{https://doi.org/10.1093/mnras/stab2843}
{Mon. Not. R. Astron. Soc. {\bf 508}, 3820--3847}
\href{https://doi.org/10.1093/mnras/stab2843}
{(2021)}.

\bibitem{FizikaKosmosa}
{\em Physics of the Space. Short Encyclopedia},
Ed. by R.\,A.\,Syunyaev (Sov. Entsikl., Moscow, 1986) [in Russian].

\bibitem{MTW}
C.\,W.\,Misner, K.\,S.\,Thorne, and J.\,A.\,Wheeler,
{\em Gravitation} (Freeman, San Francisco, 1973).

\bibitem{HaradaKimura11b}
T.\,Harada and M.\,Kimura,
{\em Collision of an object in the transition from adiabatic inspiral to
 plunge around a Kerr black hole},
\href{https://doi.org/10.1103/PhysRevD.84.124032}
{Phys. Rev. D {\bf 84}, 124032 (2011)}).

\bibitem{Nayakshin12}
S.\,Nayakshin, S.\,Sazonov, and R.\,Sunyaev,
{\em Are supermassive black holes shrouded by ‘super-Oort’ clouds of comets
and asteroids?}
\href{https://doi.org/10.1093/mnras/stab2843}
{Mon. Not. R. Astron.}
\href{https://doi.org/10.1093/mnras/stab2843}
{Soc. {\bf 419}, 1238--1247 (2012)}.

\bibitem{GribPavlov2019i}
A.\,A.\,Grib and Yu.\,V.\,Pavlov,
{\em On the limiting energy of the collision of elementary particles close to
horizon of the rotating black hole},
\href{https://doi.org/10.1142/S0217732320502624}
{Mod. Phys. Lett. A {\bf 35}},
\href{https://doi.org/10.1142/S0217732320502624}
{2050262 (2020)}.

\bibitem{Weinberg}
S. Weinberg,
{\em Gravitation and Cosmology: Principles and applications of the general
theory of relativity} (Wiley, New York, 1972).

\bibitem{GribVac}
A.\,A.\,Grib, {\em Vacuum Noninvariance Problem in Quantum Field Theory}
(Atomizdat, Moscow, 1978) [in Russian].

\end{thebibliography}
\end{document}